\documentclass[showpacs,preprintnumbers,amsmath,amssymb, prl, twocolumn]{revtex4-1}

\usepackage{stmaryrd}
\usepackage{txfonts}
\usepackage{amssymb}
\usepackage{mathrsfs}
\usepackage{graphicx}
\usepackage{dcolumn}
\usepackage{bm}
\usepackage{epsfig}
\begin{document}

\title{Massless Leggett Mode in Three-band Superconductors with Time-Reversal-Symmetry Breaking}

\author{Shi-Zeng Lin and Xiao Hu}

\affiliation{WPI Center for Materials Nanoarchitectonics, National
Institute for Materials Science, Tsukuba 305-0044, Japan
}

\begin{abstract}
The Leggett mode associated with out-of-phase oscillations of
superconducting phase in multi-band superconductors usually is heavy
due to the interband coupling, which makes its excitation and
detection difficult. We report on the existence of a massless
Leggett mode in three-band superconductors with
time-reversal-symmetry-breaking (TRSB). The mass of the Leggett mode
is small close to the TRSB transition and vanishes at the transition
point, and thus locates within the smallest superconducting energy
gap, which makes it stable and detectable. The mass of the Leggett
mode can be measured by Raman spectroscopy. The thermodynamic
consequences of this massless mode and possible realization in
iron-based superconductors are also discussed.
\end{abstract}


\date{\today}

 \maketitle
\noindent {\it Introduction --} Spontaneous breaking of a continuous
symmetry  and the associated low-energy collective excitation govern
the physical properties in many systems ranging from condensed
matter physics to particle physics. Superconductivity emerging as
the spontaneous breaking of the $U(1)$ gauge symmetry supports a
massless excitation known as Bogoliubov-Anderson-Goldstone (BAG)
boson\cite{Anderson58,Bogoliubov59}. Coupled with electromagnetic
field, the BAG boson becomes the massive plasma mode due to the
Anderson-Higgs mechanism.

\begin{figure}[t]
\psfig{figure=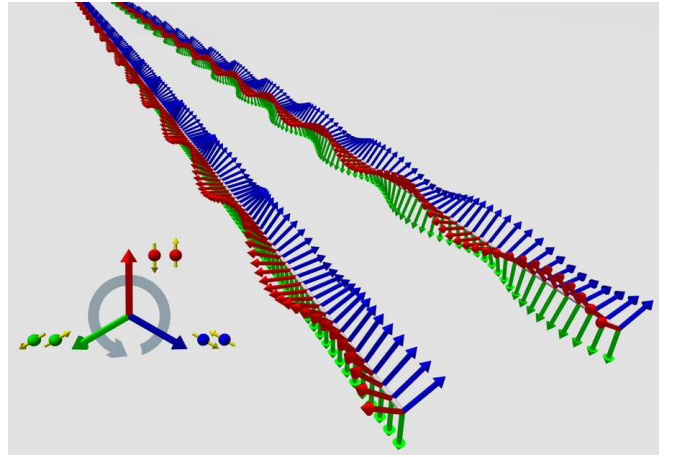,width=\columnwidth}
 \caption{\label{f0}
Frustrated interband scatterings force Cooper pairs in different
bands carry different phases, which results in interband Josephson
currents. Two dynamical modes associated with superconducting phases
in three-band superconductors: the LM, where one of the
three phases stays unchanged while the other two oscillate
out-of-phase, becomes massless at the TRSB transition (left), and
the BAG mode, where all the three phases rotate in the same
direction during the propagation of plasma wave in space (right). }
\end{figure}

Because of the discoveries of $\rm{MgB_2}$\cite{Nagamatsu01} and
iron  pnictides\cite{Kamihara08}, it is now accepted that
multi-component superconductors are ubiquitous. Multi-band
superconductors are not straightforward extensions of the
single-band counterpart, novel features arise
instead\cite{Leggett66,Agterberg99,Tanaka02,Babaev02,Blumberg07,Xi08}.
A famous example is the Leggett mode (LM) in two-band superconductors
associated with the collective oscillation of superconducting
condensates between different bands, as schematically depicted in
Fig. \ref{f0}, with the mass proportional to the interband coupling.
\cite{Leggett66} In 2007, Blumberg \emph{et. al.} reported the
observation of the LM in $\rm{MgB_2}$ with the Raman
spectroscopy\cite{Blumberg07}. The mass of the mode lies between the
two superconducting energy gaps, consistent with the theoretical
calculations\cite{Sharapov02}. The LM in $\rm{MgB_2}$
therefore decays into quasiparticle continuum associated with the
band of smaller energy gap. The heavy LM in $\rm{MgB_2}$
has also been observed in point-contact transport
measurements\cite{Ponomarev04}.

For iron-based superconductors, many studies have revealed the
sign-reversal pairing symmetry between different
bands\cite{Mazin08,Kueoki08,Ding08,Chen10}. The system of more than
three bands is somehow frustrated, and under appropriate conditions
there may exist time-reversal-symmetry-breaking (TRSB) states even
with conventional s-wave pairing symmetry, which involve non-trivial
phase differences (i.e. $\delta\varphi\neq$ 0 or $\pi$) among superconducting
gaps\cite{Agterberg99,Stanev10,Tanaka10,Hu11}. With the new TRSB
transition below $T_c$, the transition temperature of
superconductivity, the spectrum of collective excitations and thus
low-energy physical properties of the superconductors should be
modified significantly. It was reported that the LM may
exist below the two-particle continuum in iron-based superconductors
under appropriate conditions, \cite{Burnell10} and that the mass of
the LM may be reduced in some dynamical classes of
multiple interband Josephson coupling in three-band
superconductors.\cite{Ota11} We note that in another TRSB superconducting systems with
mixed-symmetry order parameters with nodes such as $d+i s$, a massive LM in the TRSB state was found in Ref. \cite{Balatsky00}. With all these recent progresses in
mind, we ask a question which is of fundamental interest: what are
the effects of frustration on the LM and is it possible to
have a massless LM mode?

In the present work, we demonstrate that the mass of the LM can be
reduced significantly and even vanishes at TRSB transition
upon sweeping interband coupling or density of states in multi-band
superconductors. It is shown that the LM of vanishing mass can be
detected by Raman scattering, which also serves as a smoking gun
evidence for the TRSB transition. The appearance of massless
excitation modifies superconductivity properties qualitatively, such
as the power-law dependence of the specific heat (SH) on temperature
instead of the conventionally exponential one for full-gapped
systems. It is found that several recent experiments on
iron-based superconductors can be explained by the existence of
massless LM.

\vspace{3mm} \noindent {\it Leggett mode --} The Hamiltonian for
three separate pieces of the isotropic Fermi  surface can be written
as
\begin{equation}\label{eq1}
\begin{array}{l}
H = \sum\limits_{l,\sigma } {\int {{d^3}r\psi _{l\sigma }^\dag (\mathbf{r}){(\varepsilon_l-\mu)} {\psi _{l\sigma }}(\mathbf{r})} } \\
 - \sum\limits_{j,l} {\int } {d^3}r\psi _{j\sigma }^\dag (\mathbf{r})\psi _{j\bar{\sigma} }^\dag (\mathbf{r}){V_{jl}}{\psi _{l \bar{\sigma} }}(\mathbf{r}){\psi _{l\sigma }}(\mathbf{r}),
\end{array}
\end{equation}
where $\psi _{l\sigma }^\dag$ (${\psi _{l\sigma }}$) is the electron
creation (annihilation) operator in the $l$-th band with the
dispersion $\varepsilon_l(\mathbf{k})$ and the chemical potential
$\mu$ and spin index $\sigma$. $V_{jl}$ is the intraband for $l=j$
and interband for $l\neq j$ scattering respectively, which can be
either repulsive or attractive depending, for instance, on
the strength of the Coulomb and electron-phonon interaction. The
interband repulsion may cause frustration of the superconductivity
in different bands and results in TRSB\cite{Agterberg99,Stanev10}.
Introducing the Nambu spinor operator $\Psi_j=(\psi_{j\uparrow},
\psi_{j\downarrow}^\dag)^T$  and the energy gap $\Delta_j$ through
the Hubbard-Stratonovich transform, we arrive at the following
action in the imaginary time representation after integrating out
the fermionic fields\cite{SimonsQFT}
\begin{equation}\label{eq2}
{S} = \int d \tau d^3 r\sum_{j,l}^3{\Delta _j}{g_{jl}}\Delta _l^* - \sum\limits_j {{\rm{Tr}}} \ln {\cal G}_j^{ - 1},
\end{equation}
with $\hat{g}=\hat{V}^{-1}$ and the Gor'kov green function
\begin{equation}\label{eq2}
{\cal G}_j^{ - 1} =  - \left( {\begin{array}{*{20}{c}}
{{\partial _\tau } + ( {{\varepsilon_j} - \mu } )}&{ - {\Delta _j}}\\
{ - \Delta _j^*}&{{\partial _\tau } - ( {{\varepsilon_j} - \mu } )}
\end{array}} \right).
\end{equation}
The superconducting energy gaps at $T=0$ are given by
\begin{equation}\label{eq2a}
\sum\limits_{l=1}^3 {{\Delta _l}{g_{{{lj}}}}}  = {N_j}(0){\Delta
_j}\sinh^{-1}\left (\frac{\hbar\omega _{cj}}{|\Delta _j|} \right),
\end{equation}
with $N_j(0)$ the density of states (DOS) at the Fermi surface in
normal state. Here $\omega _{cj}$ is a cutoff frequency and depends
on the pairing mechanism. For electron-phonon coupling,
$\omega_{cj}$ is the Debye frequency.

For demonstration of our basic idea, we take a set of simplified interband couplings\cite{supplement}
\begin{equation}\label{eqg1}
\hat g = \frac{1}{V}\left( {\begin{array}{*{20}{c}}
\alpha &1&1\\
1&\alpha &\eta \\
1&\eta &\alpha
\end{array}} \right),
\end{equation}
and assume that the DOS $N$ and $\omega_c$ are identical for three bands.\cite{Stanev10} The
massless LM, however, is not restricted to the specific
choice of $\hat g$ as discussed later. Here $g_{ij}>0$
corresponds to a repulsive interaction. We take $\Delta_1$ as
positive real, and $\Delta_2=\Delta e^{i \varphi}$, $\Delta_3=\Delta
e^{-i \varphi}$ because they are symmetric under the condition of
Eq. (\ref{eqg1}). Hereafter we take $\hbar \omega_c$ as the unit for
$\Delta_l$.

\begin{figure}[t]
\psfig{figure=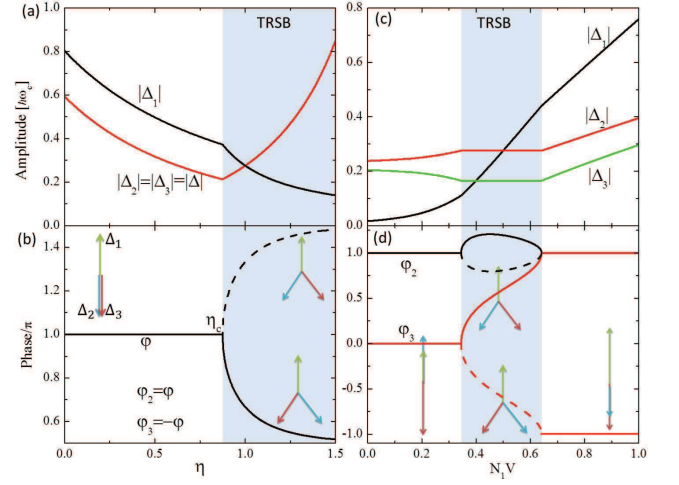,width=\columnwidth} \caption{\label{f1}
(color online). Amplitudes and phases of order parameters at TRSB
phase transition, in (a) and (b) as a function of $\eta$, and in (c,
d) as a function of DOS $N_1 V$ of the first component. $\Delta_1$
is taken as real and positive. In (a) and (b), an identical DOS $N
V=0.5$ is taken for the three bands and $\alpha=2$ in Eq.~(5). In
(c) and (d), $N_2 V=0.5$ and $N_3 V=0.4$, $\alpha=2$ and $\eta=1$
(see also Eq.~(17)). In the TRSB regime, there are two degenerate
ground states $(\Delta_1, \Delta_2, \Delta_3)$ (solid lines) and
$(\Delta_1^*, \Delta_2^*, \Delta_3^*)$ (dashed lines). The
two solid lines for $N_1 V>0.64$ in (d) refer to the same state without
TRSB.}
\end{figure}

For a small $\eta$, the interband repulsion $g_{12}$ and $g_{13}$ dominates and the system takes $\varphi=\pi$. For a large $\eta$, a state with finite phase difference between
$\Delta_2$ and $\Delta_3$ appears, corresponding to a state of TRSB
where $(\Delta_1, \Delta_2, \Delta_3)\neq (\Delta_1, \Delta_2,
\Delta_3)^*$, even apart the common phase factor. In the
TRSB state the energy gaps are given by
\begin{equation}\label{eq3}
{\Delta _1} = 1/{{\sinh\left( {\frac{{\alpha \eta  - 1}}{\eta }\frac{1}{{NV}}} \right)}} \text{,\ \ and\ \ }  \Delta  =1/{\sinh \left( {\frac{\alpha  - \eta }{N V}} \right)}
\end{equation}
and $\cos \varphi=-\Delta_1/(2\eta \Delta)$.
The system undergoes a second-order TRSB transition at $\eta_c$ given by $\eta_c=\Delta_1(\eta_c)/[2 \Delta(\eta_c)]$, as shown in Fig. \ref{f1} (a) and (b).

We proceed to investigate phase fluctuations at the TRSB point, where the amplitudes of the
superconducting gap can be considered as rigid. For this purpose, we
perform the following gauge transformation which separates the phase and amplitude of gap
\cite{Loktev01,Sharapov02}
\begin{equation}
{\Delta _j} \rightarrow {|\Delta _j|}{e^{i{\theta _j}}}{\text{\ and\ }}{\Psi _j}(\tau ,r) \rightarrow \left( {\begin{array}{*{20}{c}}
{{e^{i{\theta _j}/2}}}&0\\
0&{{e^{ - i{\theta _j}/2}}}
\end{array}} \right){\Psi _j}(\tau ,r).
\end{equation}
and derive the action for the phase fluctuation
\begin{equation}\label{eq6}
S = \int d \tau d^3r\sum_{j,l} {{|\Delta _l|}{g_{{{lj}}}}{|\Delta _j|}e^{i( {\theta _l - \theta _j})}} - \sum\limits_j {{\text{Tr}}} \left[ {\ln \left( {{\cal G}_j^{ - 1} - {\Sigma _j}} \right)} \right]
\end{equation}
where $\Sigma_j={ - \frac{{{\hbar
^2}}}{{2{m_j}}}}({\frac{i}{2}{\nabla ^2}{\theta _j}  + i\nabla
{\theta _j}\nabla })\sigma_0+[{i\frac{{{\partial _\tau }{\theta
_j}}}{2}}+{\frac{{{\hbar ^2}}}{{8{m_j}}}}{{{\left( {\nabla {\theta
_j}} \right)}^2}}]\sigma_3$ with $\sigma_j$ being the Pauli
matrices, $\sigma_0$ the unit matrix and $m_j$ the electron mass
\cite{Palo99,Benfatto04}. From this action, one can obtain the
time-dependent nonlinear Schr\"{o}dinger Lagrangian for the phase
fluctuations\cite{Aitchison95,Aitchison00}. Considering small phase
fluctuations around the saddle point $\phi_j=\theta_j-\varphi_j$ and
expanding $S$ up to the second order in $\phi_j$, we
have\cite{supplement}
\begin{equation}\label{eq7}
S_\phi\left[ {{\phi _j}} \right] = \frac{1}{8}\sum\limits_l \int{{d^3}}q \hat{\phi}({-\Omega _l},-q)^T {\bf{M}}
\hat{\phi}({\Omega _l},q)
\end{equation}
with $\hat{\phi}({\Omega _l},q)\equiv [\phi _1({\Omega _l},q), \phi _2({\Omega _l},q), \phi _3({\Omega _l},q)]^T$ and
\begin{equation}\label{eq8}
{\bf{M}} = \left( {\begin{array}{*{20}{c}}
{ P_1 - 2{D_1}}&{{D_1}}&{{D_1}}\\
{{D_1}}&{ P_2- {D_1} - {D_2}}&{{D_2}}\\
{{D_1}}&{{D_2}}&{ P_3 - {D_1} - {D_2}}
\end{array}} \right)
\end{equation}
with ${D_1} = {8}{\Delta _1}\Delta \cos{{\bar{\varphi}}}/V$ and ${D_2} = {{8\eta }}{\Delta ^2}\cos ( {2{\bar{\varphi}}} )/V$ with $\bar{\varphi}\equiv\varphi_2-\varphi_1=\varphi_1-\varphi_3$. $\Omega_l=2l\pi k_B T $ and the excitations are bosons. In the hydrodynamic limit at $T=0$, the dissipation is absent and $P_j={ 2N(-\Omega^2 +  {{1}/{3}v_j^2} {q^2})}$ after the analytical continuation $i\Omega_l\leftarrow\Omega+i 0^+$. From $\text{Det}\mathbf{M} =0$, we obtain the dispersion relations
\begin{eqnarray}
  \Omega _{\text{BAG}}^2 &=& \frac{1}{3}{q^2}v_j^2, \\
  \Omega _{\text{L-}}^2 &=& {-\frac{ {D_1} + {2D_2}}{2N} + \frac{1}{3}{q^2}v_j^2}, \\
  \Omega _{\text{L+}}^2 &=&  { - \frac{{3D_1}}{2N} +  \frac{1}{3}{q^2}v_j^2} .
\end{eqnarray}

\begin{figure}[t]
\psfig{figure=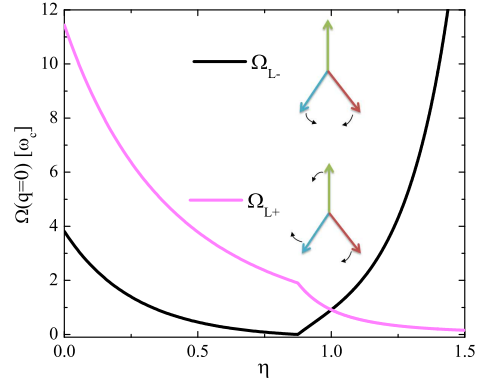,width=7.1cm} \caption{\label{f2} (color online). Dependence of the
masses of LMs on the interband coupling $\eta$. Here $N V=0.5$ and $\alpha=2$, and the masses are
in units of $\omega_c$.}
\end{figure}

The first mode is the massless BAG mode corresponding to the uniform rotation of phases. The second and third are the LM $\Omega_\text{L-}$ and $\Omega_\text{L+}$ in the present three-band system. Especially, the mode $\Omega_\text{L-}$
corresponds to the dynamics of the relative phase $\varphi_{23}$ between the gaps of $\Delta_2$ and $\Delta_3$, and becomes massless
at the TRSB transition depicted in Fig. \ref{f2}.
One may regard $\varphi_{23}$ as the order parameter for the TRSB transition. As it increases continuously from $0$ at the transition, the associated fluctuations become massless at the TRSB transition.

The magnetic field can be introduced into $S_{\phi}$ through the standard replacement
$\nabla\phi_l\rightarrow\nabla\phi_l -2\pi \mathbf{A}/\Phi_0$ with $\Phi_0$ the flux quantum and $\mathbf{A}$ the vector potential. In this case, it is more convenient to
describe the phase fluctuations in terms of $\phi_1$, $\phi_{12}\equiv\phi_1-\phi_2$ and
$\phi_{13}\equiv\phi_1-\phi_3$. $\phi_1$ describes the BAG mode, and $\phi_{12}$ and $\phi_{13}$ correspond to
the LMs. The gauge field couples with $\phi_1$ in the form $(\nabla\phi_1-2\pi \mathbf{A}/\Phi_0)$. One may integrate
out $\phi_1$, resulting in the massive plasma mode due to the Anderson-Higgs mechanism. In contrast to the
BAG mode, the LMs remain massless at the TRSB transition since $\phi_{12}$ and $\phi_{13}$ are
decoupled from the gauge field $\mathbf{A}$.

In stark contrast to conventional symmetry-broken systems, there
exist the stable LMs both before and after TRSB transition, because
the relative phase between different condensates is fixed in both
the states with and without TRSB.

\vspace{3mm} \noindent {\it Raman scattering--} Interband
scatterings do not involve the gauge field, thus the LMs do not
respond to a magnetic field. However, the LMs are coupled indirectly
with the electric field through the charge density, which renders it
detectable by the Raman spectroscopy through the inelastic
scattering of photon with the charge
density\cite{Abrikosov61,Klein84,Devereaux95,Lee09}. The interaction
between the incident photon and the charge can be modeled as $\tilde
\rho (\tau ,q) = \sum\limits_{j = 1}^3 {\sum\limits_{k,\sigma }
{{\gamma_j }} } (k)\psi _{{{j}}\sigma }^\dag \left( {\tau ,k +
\frac{q}{2}} \right){\psi _{{{j}}\sigma}}\left( {\tau ,k -
\frac{q}{2}} \right)$, where $\gamma_j(k)$ is the scattering
coefficient determined by the polarization of the incident and
scattered photon. In the following, we derive the experimentally
measurable Raman response function ${{\chi _{\tilde \rho \tilde \rho
}}(\tau  - \tau ',q) =  - \left\langle {{\text{T}_\tau }\tilde \rho
(\tau ,q)\tilde \rho (\tau ', - q)} \right\rangle }$ with
$\text{T}_\tau$ being time-ordering operator. We introduce a source
term coupled with $\tilde \rho$, ${H_J}(\tau ) =  - \sum\limits_q
{{{\tilde \rho }}} (\tau, q){J}(\tau, -q )$ because $\chi _{\tilde
\rho \tilde \rho }$ can be computed by the linear response theory
with respect to $J$. The effective action in the presence of
incident photon reads\cite{supplement}
\begin{equation}
S = \int d \tau d^3r\sum\limits_{l,j} {{\Delta _l}} {g_{{lj}}}\Delta _j^* - \sum\limits_l {{\rm{Tr}}} \ln \left( {{\cal
G}_{J,l}^{ - 1} + {\cal G}_{l}^{ - 1}} \right)
\end{equation}
with ${\cal G}_{J,l}^{ - 1} =  - {\gamma _l}(k)J(\tau , - q){\sigma _3}$. For a weak incident wave, we may neglect the
fluctuations of the amplitude of the order parameters, and the fluctuations for the superconducting phase acquires a form ${S } =S_{\phi}+S_{\text{J}}$,
with $S_{\phi}$ defined in Eq. (\ref{eq7}) and
\begin{equation}\label{eq11b}
\begin{array}{l}
S_{\text{J}}= \frac{1}{2}\sum\limits_{j,q}[J(q){Z_j}(q)\phi_j^T( - q) \\
+ J( - q){{\tilde Z}_j}( - q)\phi _j(q)+ J (q)J( - q)\Pi _{j,33}^{\gamma \gamma }],
\end{array}
\end{equation}
where ${Z_j}(q) = \Delta_j[-\sin \varphi _j\Pi _{j,31}^\gamma (q) - \cos\varphi_j \Pi _{j,32}^\gamma (q) ]$ and
${{\tilde Z}_j}(q) = \Delta_j[ -\sin \varphi _j\Pi _{j,13}^\gamma (q) - \cos\varphi _j \Pi _{j,23}^\gamma (q) ]$. The
polarization functions are defined as $\left[\Pi_{j,ml}^{\gamma \gamma}, \Pi_{j,ml}^{\gamma }\right]\equiv
{1}/(L^3\beta )\sum _n\int d^3k\Upsilon_{j, ml}\left[\gamma_j(k+\frac{q}{2})\gamma_j(k-\frac{q}{2}),
\gamma_j(k+\frac{q}{2})\right]$ .

\begin{figure}[t]
\psfig{figure=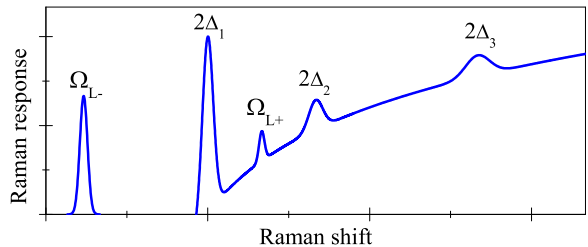,width=\columnwidth} \caption{\label{f3} (color online). Schematic view of the Raman response in three-band superconductors with TRSB. The finite line-width of peaks is due to damping and interaction between Leggett bosons. The background at energy larger than $2\Delta_1$ is due to the quasiparticle excitations.}
\end{figure}

Integrating out the fluctuations $\phi_j$, we then obtain the correlation function
\begin{equation}\label{eq12}
{{\chi _{\tilde \rho \tilde \rho }}(i\Omega ,q = 0)}=\sum\limits_j \left\{{\Pi _{j,33}^{\gamma \gamma }}  - {Z_j} [\mathbf{M} ^{ - 1}]_{jj}{ {{{\tilde Z}_j^T}} }\right\}.
\end{equation}
The first term gives the resonant scattering at $\Omega=2\Delta_j$
and  the second term accounts for the resonance with the LMs, as depicted in Fig. \ref{f3}. When the energy shift of the
photon matches the energy of the LMs, $\mathbf{M} ^{ - 1}$
becomes singular and gives $\delta$ peaks in the spectroscopy. In
reality, the delta-function peaks are rounded by both damping effect
and interactions between Leggett bosons when the oscillations of the
LM become strong, which are absent in Eq. (\ref{eq12}).
Although the response of a genuinely massless LM
is hidden into the elastic scatterings, it can be traced out clearly
if one changes $\eta$ systematically and generates LM of
small mass, which can be achieved by carrier doping because the
interband scattering is renormalized by the DOS as in
Eq. (\ref{eq2a}).

\vspace{3mm} \noindent {\it Discussions--} At $T>0$, the Landau
damping by quasiparticles sets in and the lifetime of the LM
decreases. A local time-dependent equation for the phase
fluctuations does not exist due to the singularity of the DOS in the
superconducting state\cite{Abrahams66}. In the vicinity of $T_c$,
the dynamics of superconductivity can be described by the standard
time-dependent Ginzburg-Landau equation\cite{Abrahams66}. In this
region, the lifetime of the LM is much smaller than the inverse of
its energy due to the severe damping by quasiparticles, therefore
there is no well-defined Leggett excitations. Nevertheless in the
static case, the massless feature manifests as the divergence of the
characteristic length for the relative phase variation in the
vicinity of the TRSB \cite{Hu11}.

Let us discuss the applicability of our results to the iron-based
superconductors. In order to demonstrate the mass reduction of the
LM by TRSB, we adopt a simple and general BCS-like Hamiltonian in
Eq.~(\ref{eq1}). An implicit expectation behind this treatment is
that more realistic models would merely lead to quantitative
corrections. It seems that this simplification is not far from the
situation in some iron-based superconductors, since the $s$-wave
with sign-reversal ($s\pm$) pairing symmetry is favored by many
experiments\cite{Ding08,Chen10}. It also became clear recently that
the interband hopping in single particle channels using a more
realistic tight-binding model gives additional contribution to the
interband Josephson coupling, and that the Hunds interaction only
gives higher order correction to the LM\citep{Burnell10}. It was
also shown\cite{Graser09} that the $s\pm$ pairing can result from
the moderate electronic
correlations\cite{Qazilbash09,Haule09,Aichhorn80,Si09,Hansmann10} in
iron-based superconductors, thus electronic correlations probably do
not hamper much the massless Leggett mode.

The reason that no direct experimental observation on the TRSB state in
iron-pnictide superconductors has been reported to date may come
from its requirement on sufficiently strong frustration interactions
among different bands. Here we wish to observe that the TRSB
transition can be induced not only by interband coupling but also by
DOS $N_j(0)$. In order to demonstrate this we derive the TRSB
solution to Eq.~(\ref{eq2a}) under a general coupling matrix $\hat
g$. Complex gap functions as solution to Eq.~(\ref{eq2a}) appear
when there is only one independent vector in the matrix $\hat g -
\hat g'$, with $g'_{jj}=N_j(0)\sinh^{-1}(\hslash\omega_{
cj}/|\Delta_j|)$ and 0 otherwise. From this constraint we
obtain\cite{supplement}
\begin{equation}
\frac{|\Delta_j|}{\hslash\omega_{
cj}}=\frac{1}{\sinh[(g_{jj}g_{kl}-g_{jk}g_{jl})/N_j(0)g_{kl}]},
\label{deltas}
\end{equation}
with $j\neq k\neq l$. It is easy to see that to find further the
phases of the gap functions is equivalent to forming a triangle with
the three segments $|\Delta_j|/g_{kl}$, which is possible when and
only when $|\Delta_j|/g_{kl}+ |\Delta_k|/g_{jl}>|\Delta_l|/g_{jk}$
for all the three combinations. The phase transition from a TRSB
state to a state without TRSB takes place when one of the above
inequalities is broken, for example $|\Delta_1|/g_{23}=
|\Delta_2|/g_{13}+|\Delta_3|/g_{12}$. The results for
DOS-driven TRSB transition are displayed in Fig. \ref{f1}
(c) and (d). There are two TRSB transitions and the TRSB state is
realized in a finite region of DOS. Therefore, experimentally one
can tune $N_j(0)$ by careful chemical doping, which hopefully will
drive the system to the TRSB transition.

Although the massless LM does not change magnetic properties of the
system, it results in qualitatively different thermodynamic
behaviors of $s$-wave superconductivity. For the SH, the
contribution due to quasiparticles at $T\ll T_c$ depends
exponentially on temperature $(\Delta/k_B T)^{3/2} \exp (-\Delta/k_B
T)$ for fully gapped superconductors. The contribution of the
massless Leggett excitations can be obtained analytically by
treating the Leggett bosons as free quantum gas. The contribution is
of power-law temperature dependence $T^3$, which can be detected
experimentally.

It is worth noting that a $T^3$ dependence of the SH in
iron-base superconductors after subtracting the residue electronic
contribution (linear in $T$) and phonon contribution (also $T^3$
dependence) has been reported in several experiments,
\cite{Kim10,Gofryk83,Zeng11}; fully gapped order parameters are
inferred from measurements for the dependence of electronic SH on
magnetic field, which excluded the possibility of gap function of
line node. Actually, in Ref. \cite{Gofryk83}, the authors suggested
that the additional $T^3$ contribution might be due to some bosonic
modes. These experimental observations can be naturally explained by
the existence massless LM. Additional measurements such as the Raman
spectroscopy on similar samples \cite{Kim10,Gofryk83,Zeng11} are
much anticipated which may well be in the vicinity of the TRSB
transition.

\vspace{3mm}\noindent {\it Acknowledgements --} The authors are
grateful for L. Bulaevskii, J.~-X. Zhu and Z. Wang for discussions.
This work was supported by WPI Initiative on Materials
Nanoarchitectonics, and Grants-in-Aid for Scientific Research
(No.22540377), MEXT, Japan, and partially by CREST, JST.

%

\end{document}